\documentclass{article}
\usepackage{spconf,amsmath,graphicx,hyperref,xcolor}
\usepackage{subfigure}
\usepackage{tabu}                     
\usepackage{multicol}                 
\usepackage{multirow}                 
\usepackage{float}                    
\usepackage{makecell}                 
\usepackage{booktabs}                 
\usepackage{tabularray}

\usepackage{enumitem}
\setlist{nosep, leftmargin=14pt}

\usepackage{mwe} 

\title{Optimized Hard Exudate Detection with Supervised Contrastive Learning}
\name{\begin{tabular}{c} Wei Tang $^{1,2}$, Kangning Cui $^{1,2}$\sthanks{Corresponding Author.}, Raymond H. Chan $^{1,2}$\end{tabular}}
\address{$^1$ Hong Kong Centre for Cerebro-Cardiovascular Health Engineering, Hong Kong\\
$^2$ Department of Mathematics, City University of Hong Kong, Hong Kong}

\begin{document}
\maketitle
\begin{abstract}
Diabetic retinopathy (DR) is a leading global cause of blindness. Early detection of hard exudates plays a crucial role in identifying DR, which aids in treating diabetes and preventing vision loss. However, the unique characteristics of hard exudates, ranging from their inconsistent shapes to indistinct boundaries, pose significant challenges to existing segmentation techniques. To address these issues, we present a novel supervised contrastive learning framework to optimize hard exudate segmentation. Specifically, we introduce a patch-wise density contrasting scheme to distinguish between areas with varying lesion concentrations, and therefore improve the model's proficiency in segmenting small lesions. To handle the ambiguous boundaries, we develop a discriminative edge inspection module to dynamically analyze the pixels that lie around the boundaries and accurately delineate the exudates. Upon evaluation using the IDRiD dataset and comparison with state-of-the-art frameworks, our method exhibits its effectiveness and shows potential for computer-assisted hard exudate detection. The code to replicate experiments is available at \href{https://github.com/wetang7/HECL/}{github.com/wetang7/HECL/}.
\end{abstract}


\section{Introduction}
\label{sec:intro}


Fundus images, captured using a specialized fundus camera, provide detailed views of the interior surface of the eye, including the retina, blood vessels, and other structures~\cite{alyoubi2020diabetic, asiri2019deep}. A pivotal application of these images is in the detection of diabetic retinopathy (DR), a complication of diabetes that affects the eyes. Among the earliest clinical signs of DR, hard exudates are notable since they are closely associated with vision damage in the early stages of DR~\cite{sanchez2009retinal}. Addressing them promptly is vital not only to prevent vision loss but also due to the increased risk of cardiovascular diseases (CVDs) associated with DR~\cite{cheung2008diabetic}. Deep learning techniques, in particular, have excelled in medical image applications, thus aiding consistent diagnosis~\cite{ronneberger2015u, zhou2018unet++, pan2023dual}. To tackle the challenges of hard exudate detection, deep learning algorithms offer a more efficient and consistent alternative to manual labeling, which is labor-intensive and prone to errors~\cite{asiri2019deep, si2021hard}. 

Despite the efficacy of existing networks, current hard exudate segmentation methods grapple with limitations due to the fine-grained, irregular shapes and non-uniform distribution of exudates across fundus images that complicate the segmentation~\cite{si2021hard, mo2018exudate}. Furthermore, these exudates frequently exhibit indistinct boundaries that are neither clear nor well-defined, leading to challenges in segmentation and potentially suboptimal results~\cite{sanchez2009retinal, guo2020bin}. Addressing these challenges, supervised contrastive learning proves effective by utilizing label information to distinctly separate different classes~\cite{khosla2020supervised}, therefore managing varying lesion densities and unclear boundaries. This motivates our framework that applies supervised contrastive learning to improve hard exudate segmentation.

Our main contributions are highlighted as follows: (1) We propose a patch-wise density contrasting scheme that contrasts regions with varying lesion concentrations, which enhances the model's ability to distinguish between lesion-dense and lesion-sparse patches. (2) We design a discriminative edge inspection module using morphological operations to precisely define and analyze lesion boundaries. (3) Extensive experiments on the IDRiD dataset demonstrate our framework's effectiveness, outperforming state-of-the-art models and performing well across various backbones, thereby confirming its robustness and adaptability.

\section{Related Works}
\label{sec:format}

\subsection{Medical Image Segmentation}

Deep learning has significantly enhanced medical image segmentation that helps with diagnosis~\cite{ronneberger2015u, zhou2018unet++, gu2019net, sang2021super, oktay2018attention, li2018h}. U-Net stands out for its integration of both a contracting and expansive path, effectively combining structure and context~\cite{ronneberger2015u}. UNet++ introduces nested blocks and deep supervision to improve segmentation accuracy~\cite{zhou2018unet++}while CE-Net counters U-Net's potential spatial detail loss with dense atrous convolution and multi-kernel pooling~\cite{gu2019net}. The CogSeg network refines segmentation through improved resolution and edge detection~\cite{sang2021super}. Attention U-Net emphasizes target structures using attention gates~\cite{oktay2018attention}, and H-DenseUNet combines 2D and 3D networks for comprehensive segmentation~\cite{li2018h}. These methods are designed for general medical image segmentation or specific applications like CT images, yet none are tailored for hard exudates with their distinct lesion structures.

\subsection{Hard Exudate Detection}
\begin{figure}[t]
    \centering
    \includegraphics[width = 0.85\linewidth]{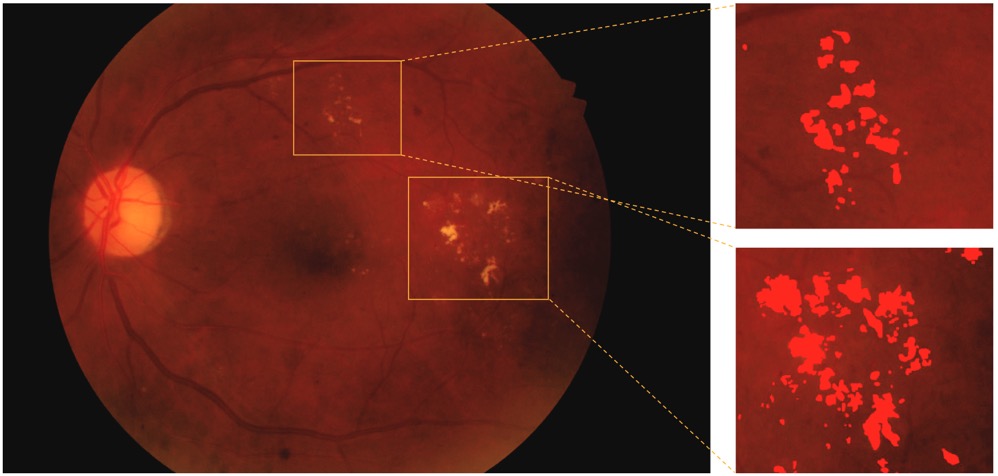}
    \vspace{-0.3cm}
    \caption{Illustration of hard exudates in a fundus image. The red pixels in the right image locate the hard exudate lesions.}
    \label{fig:he}
    \vspace{-0.5cm}
\end{figure}

Hard exudate is a common clinical symptom of diabetic retinopathy, characterized by irregular white or yellowish-white accumulations in the retina resulting from plasma leakage (refer to Figure~\ref{fig:he})~\cite{alyoubi2020diabetic, asiri2019deep}. Manifesting as dots, patches, or circles, they serve as primary indicators of potential blindness, underscoring the critical importance of early detection and treatment~\cite{sanchez2009retinal, si2021hard}. With the rapid advancement of deep learning techniques, several novel models have emerged for hard exudate segmentation~\cite{sanchez2009retinal, si2021hard, mo2018exudate, guo2020bin, guo2019exudate, zhang2022hard}. CNN-based models, in particular, utilize multi-level hierarchical information for precise segmentation~\cite{si2021hard, mo2018exudate, guo2019exudate, zhang2022hard}. In parallel, diverse loss functions have been proposed to optimize these networks and mitigate class imbalance~\cite{mo2018exudate, guo2020bin, xie2015holistically}. While substantial progress has been made, challenges such as the dispersion of tiny lesions and indistinct boundaries persist, necessitating continued refinement for enhanced accuracy and reliability of hard exudate detection methods.

\subsection{Contrastive Learning}

Contrastive learning, a self-supervised technique, learns data representations by contrasting positive (similar) and negative (dissimilar) sample pairs, mapping them closely or distantly in feature spaces~\cite{chen2020simple, camalan2022detecting}. While unsupervised contrastive learning has shown promise in various applications, its feature representations might not be task-specific, given the absence of supervision~\cite{khosla2020supervised}. Supervised contrastive learning incorporates labels, contrasting samples based on labels rather than mere data augmentation. This method optimizes the loss function to enhance the distinction between positive and negative pairs, yielding more discriminative features suitable for specific tasks like hard exudate detection~\cite{khosla2020supervised, islam2022applying}.

Contrastive learning, while proven for classification tasks such as DR grading, is yet to be fully explored for segmentation in DR~\cite{islam2022applying}. Furthermore, the adoption of multi-level feature approaches in contrastive learning shows promise in enhancing segmentation performance~\cite{wu2022cross, zhao2022cross}. The gap in research presents an opportunity for advancements in medical image analysis, specifically in the segmentation of hard exudates, which could benefit from the detailed feature discrimination that contrastive learning provides.
\section{Methodology}
\label{sec:Methodology}

\begin{figure}[t]
    \centering
    \includegraphics[width = \linewidth]{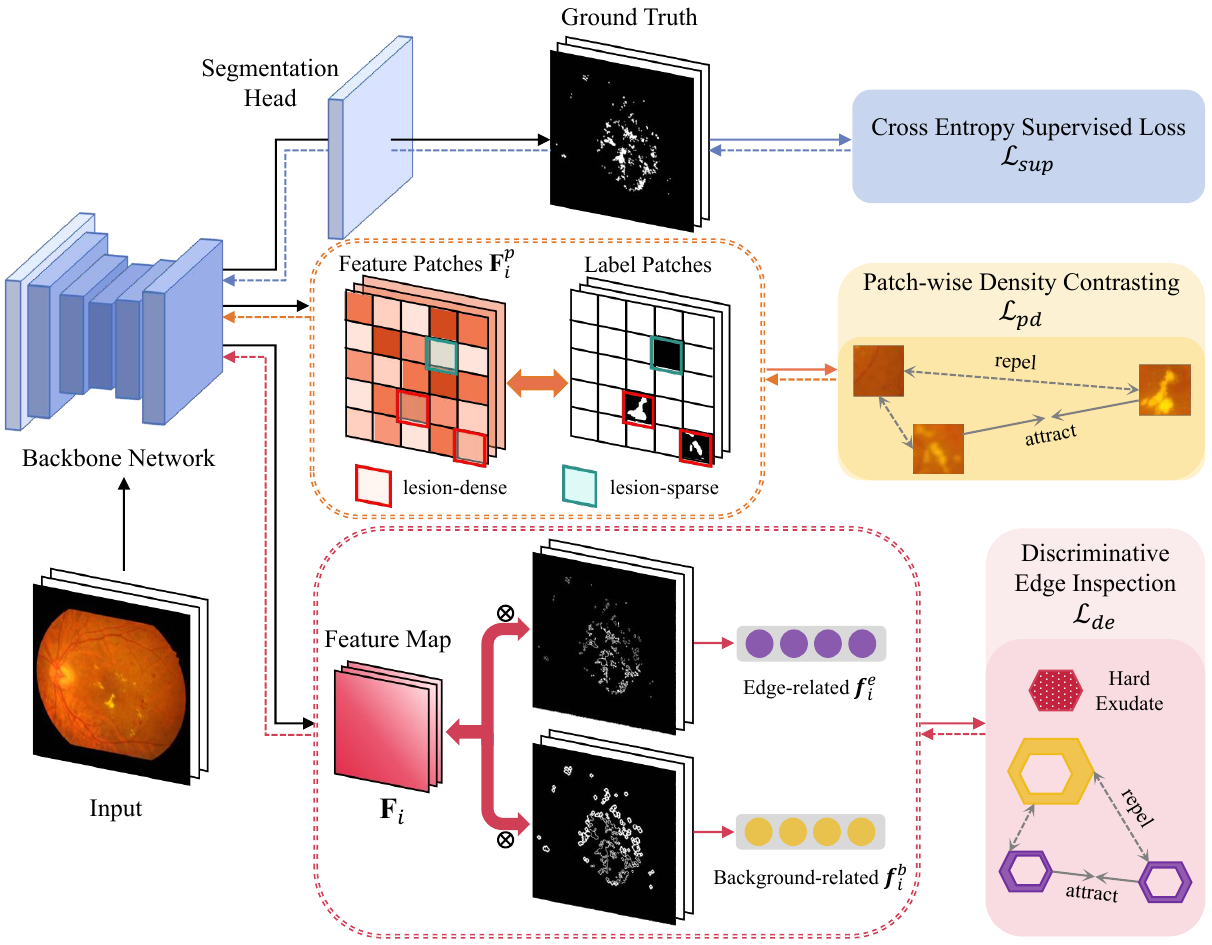}
    \vspace{-0.5cm}
    \caption{An overview of the proposed framework. The network is jointly trained by $\mathcal{L}_{sup}$, $\mathcal{L}_{pd}$, and $\mathcal{L}_{pe}$ in order to learn both ``density-aware'' and ``boundary-aware'' knowledge.}
    \label{fig:pipeline}
    \vspace{-0.5cm}
\end{figure}

Given a fundus image dataset with \(N\) samples, denoted as $\mathcal{D} = \{(\boldsymbol{x}_i, \boldsymbol{y}_i) \mid i = 1, \ldots, N\}$, each data pair consists of an input RGB image $\boldsymbol{x}_i$ and its associated mask of hard exudates $\boldsymbol{y}_i$. Our objective is to train a model that generates a pixel-wise segmentation map $\hat{\boldsymbol{y}}_i$ for each $\boldsymbol{x}_i$, ensuring the minimization of discrepancies between $\hat{\boldsymbol{y}}_i$ and $\boldsymbol{y}_i$. Our proposed framework, depicted in Figure \ref{fig:pipeline}, integrates three key components: supervised training ($\mathcal{L}_{sup}$), patch-wise dense contrasting ($\mathcal{L}_{pd}$), and discriminative boundary inspection ($\mathcal{L}_{pe}$). The backbone network is refined by jointly optimizing the loss:
\begin{equation}
\mathcal{L}_{total}=\mathcal{L}_{sup}+\alpha(\mathcal{L}_{pd}+ \mathcal{L}_{de}).
\end{equation}
The $\alpha$ is a hyper-parameter introduced to balance the weightings of distinct terms. The following subsections provide the motivations and detailed constructions of each component.

\subsection{Patch-wise Density Contrasting}
To tackle the issue of highly dispersed foreground pixel distribution and generate a more precise prediction for those fine-grained lesions, we design a patch-wise density contrasting scheme that learns from lesion pixel allocation. This approach aims to sufficiently contrast localized representations of similar and dissimilar pairs of regions, compelling the model to distinguish between the background and small clusters of hard exudates. For each input pair $(\boldsymbol{x}_i, \boldsymbol{y}_i)$, we divide the raw image and the label into $n{\times}n$ patches denoted by $\boldsymbol{x}_i^p$ and $\boldsymbol{y}_i^p$, where $p=1,...,n^2$. Each patch exhibits a varied distribution of hard exudate pixels, and we categorize all the $\boldsymbol{x}_i^p$'s into two sets---lesion-sparse and lesion-dense patches---by applying a proportion threshold of $0.3$ for lesion pixels. Regions that differ in label distributions, i.e., lesion-sparse and lesion-dense patches, should have distinct representations in the desired feature space, thereby automatically forming negative pairs. Given a feature patch $\mathbf{F}_i^p$ of size $C\times h\times w$, we regularize it into a one-dimensional vector $\boldsymbol{f}_i^p$ of size $C\times 1$:
\begin{equation}
\label{eq:1}
f_{i,c}^p = \frac{\sum_{j=1}^{h}\sum_{k=1}^{w} F_{i,c,j,k}^p}{h \times w}, \quad c = 1,2,\dots,C.
\end{equation}
Here, $F_{i,c,j,k}^p$ is the value at channel $c$, height $j$, and width $k$ of the feature map for patch $\boldsymbol{x}_i^p$. The sum is averaged over spatial dimensions for each channel. After obtaining $\boldsymbol{f}_i^p$'s, we exploit the supervised contrastive loss to enhance the model's capacity to distinguish small patches of hard exudates from the background. The patch-wise density contrastive loss $\mathcal{L}_{pd}$ for an entire mini-batch is formulated as:
\begin{equation}
\mathcal{L}_{pd}=\frac{1}{|\mathcal{M}|}\sum_{\boldsymbol{f}_i^p \in \mathcal{M}}\mathcal{L}({\boldsymbol{f}_i^p}, \mathcal{P}(p), \mathcal{N}(p)),
\end{equation}
with $\mathcal{L}({\boldsymbol{f}_i^p}, \mathcal{P}(p), \mathcal{N}(p))=$
\begin{equation}
\begin{split}
 \frac{-1}{|\mathcal{P}(p)|}\sum\limits_{\boldsymbol{f}_i^q\in \mathcal{P}(p)} \rm{log}
 \frac{ \rm{exp}( \rm{sim}(\boldsymbol{f}_i^q, \boldsymbol{f}_i^p)/\tau)}{\sum\limits_{\boldsymbol{f}_i^k\in \mathcal{P}(p)\cup \mathcal{N}(p)}\rm{exp}( \rm{sim}(\boldsymbol{f}_i^k, \boldsymbol{f}_i^p)/{\tau})},
\end{split}
\end{equation}
where $|\cdot|$ counts the cardinality of a set, $\mathcal{P}(p)$ denotes the set of positives that excludes $\boldsymbol{f}_i^p$, $\mathcal{N}(p)$ is the corresponding negative set, and $\mathcal{M}$ represents the set of stored representative features in every mini-batch. The cosine similarity function $\rm{sim}(\cdot, \cdot)$ is applied here to measure the similarity between two feature vectors, and $\tau$ stands for the temperature. 

\subsection{Discriminative Edge Inspection}
Failure to identify the blurry boundaries and differentiate them from the surrounding tissue is another contributor to the degradation of segmentation performance. To address this, we introduce a discriminative edge inspection module to dynamically analyze the pixels around the lesion boundaries and then correctly separate them into the corresponding groups.
For given patches $(\boldsymbol{x}_i^p, \boldsymbol{y}_i^p)$, morphological operations are utilized to extract the inner and outer contour masks of each patch's binary ground truth $\boldsymbol{y}_i^p$. The inner contour, containing only lesion pixels within the edge, is derived by subtracting the eroded label map from the original mask $\boldsymbol{y}_i^p$. The outer contour is obtained by subtracting $\boldsymbol{y}_i^p$ from the dilated one. Specifically, for lesion-dense patches, we set the iterations for dilation and erosion to two. For lesion-sparse patches, which typically have scattered pathological spots and a vast background, the iterations are set to five for dilation and one for erosion to gain extra information about the background. We compose the patches together to form the final inner contour $\boldsymbol{y}_{in}$ and outer contour $\boldsymbol{y}_{out}$ after the morphological operations, and we compute the corresponding averaged edge-related feature $\boldsymbol{f}_i^{e}$ and background-related feature $\boldsymbol{f}_i^{b}$ inspired by~\cite{liu2021feddg}:
\begin{equation}
\boldsymbol{f}_i^{e} = \frac{\sum\limits_{j=1}^{H} \sum\limits_{k=1}^{W} (\mathbf{F}_i \otimes \boldsymbol{y}_{in})_{jk}}{\sum_{j=1}^{H} \sum_{k=1}^{W} (\boldsymbol{y}_{in})_{jk}}, \quad
\boldsymbol{f}_i^{b} = \frac{\sum\limits_{j=1}^{H} \sum\limits_{k=1}^{W} (\mathbf{F}_i \otimes \boldsymbol{y}_{out})_{jk}}{\sum_{j=1}^{H} \sum_{k=1}^{W} (\boldsymbol{y}_{out})_{jk}}
\end{equation}
where $\otimes$ represents element-wise multiplication, $\mathbf{F}_i$ is the corresponding feature map, $H$ and $W$ are the height and width. Finally, the loss $\mathcal{L}_{de}$ is defined as:
\begin{equation}
\mathcal{L}_{de}=\frac{1}{2|\mathcal{B}|}\sum_{\boldsymbol{f}^t \in \{\boldsymbol{f}_i^{e}, \boldsymbol{f}_i^{b}\}}\mathcal{L}({\boldsymbol{f}^t}, \mathcal{P}(t), \mathcal{N}(t)).
\end{equation}
Here, $\mathcal{B}$ denotes the images in each batch. If $\boldsymbol{f}^t$ is an edge-related feature, then $\mathcal{P}(t)$ is the set of other edge-related features, and $\mathcal{N}(t)$ designates all the background-related features in a mini-batch, and vice versa. With this module, the model is expected to learn more informative representations of the pixels around the lesion boundary, and therefore develop the necessary ability to recognize the ambiguous edges.

\section{Experiments}
\label{sec:typestyle}

We evaluate our proposed method on a benchmark dataset: the Indian diabetic retinopathy image dataset (IDRiD) \cite{porwal2018indian} due to its high-quality annotations. Specifically, we use the subset annotated for hard exudates. This subset comprises 81 color fundus JPEG images of resolution 4288$\times$2848, with 54 officially designated for training. Given the dataset's limited size, we employ data augmentation techniques such as random horizontal flipping, rotation (ranging from -180$^{\circ}$ to +180$^{\circ}$), and adjustments in brightness and contrast (scaling between 50\% and 150\%) to mitigate overfitting. For model input, all images are resized and cropped to 512$\times$512 pixels.

\subsection{Experimental Setup}

We employ UNet++~\cite{zhou2018unet++} as our base network due to its prevalence in medical image segmentation tasks. The model is optimized using Adam optimizer with a batch size of 6. The initial learning rate is set as $3\times10^{-4}$ and decays by 0.5 after every 100 epochs. For the hyper-parameters, images are divided into $5\times 5$ patches, and the temperature parameter $\tau$ is set as 0.05. The balancing parameter $\alpha$ is empirically set as 0.01. To guarantee model convergence, training persists for 300 epochs for all models. The experiments are executed on an NVIDIA GeForce RTX 3090 GPU using PyTorch~\cite{paszke2019pytorch}.

For comparative analysis, we compare our approach against state-of-the-art networks including UNet++\cite{zhou2018unet++}, H-DenseUNet\cite{li2018h}, and Att-UNet~\cite{oktay2018attention} that we implemented, as well as DeepLabv3+\cite{chen2018encoder}, HED\cite{xie2015holistically}, and CogSeg~\cite{sang2021super}, whose results we directly adopt from~\cite{zhang2022hard}. Additionally, extensive ablation studies are conducted, assessing the impact of different components in the proposed loss and the robustness adaptability of our framework with alternative backbones like U-Net~\cite{ronneberger2015u}, ResUnet~\cite{zhang2018road}, and CE-Net~\cite{gu2019net}. We repeat the experiment 5 times to obtain the mean and standard deviation.

\subsection{Results and Discussion}

\begin{table}[tb]
\centering
\resizebox{0.65\linewidth}{!}{%
\begin{tabular}{l|cccc}
\toprule[1.2pt]
\textbf{Model}       & \textbf{IoU} & \textbf{F$_{1}$ score} & \textbf{AUC} & \textbf{Recall} \\ \hline
   H-DenseUNet                        & 44.59 & 59.71 & \textbf{98.50} & 52.13  \\
   Att-UNet                           & 51.81 & 66.75 & 95.67 & 59.66  \\
   DeepLabv3+                         & 41.10 & 58.21 & 84.07 & 47.81  \\
   HED                                & 52.68 & \underline{68.96} & 38.98 & \underline{65.50}   \\ 
   CogSeg                             & 46.63 & 63.44 & 63.36 & 55.56  \\ 
   \hline
   Ours                               & \textbf{55.69} & \textbf{69.64} & 96.59 & \textbf{65.65}  \\
\bottomrule[1.2pt]
\end{tabular}
}
\caption{Comparative performance of our method with state-of-the-art networks on hard exudates segmentation. Bold values indicate the best performance for each metric, while underlined values denote the second-best. For a fair comparison, we list the best performance of each model.}
\label{tab:compare}
\end{table}

The comparative results of the proposed framework and other networks are shown in Table~\ref{tab:compare}, measured in four metrics: IoU, F$_1$ score, AUC, and recall. Our framework performs the best in IoU, F$_1$ score, and recall. H-DenseUNet, despite its exceptional AUC, falters in other metrics; meanwhile, HED, with a promising recall and F$_1$ score, struggles due to a notably low AUC, both suggesting insufficient segmentation. Moreover, the inclusion of either $\mathcal{L}_{pd}$ or $\mathcal{L}_{de}$ enhances certain metrics relative to the UNet++ backbone, as shown in Table~\ref{tab:ablation}. Yet, the fusion of both $\mathcal{L}_{pd}$ and $\mathcal{L}_{de}$ in our proposed approach yields the best results with only a marginal AUC trade-off.

\begin{table}[tb] 
\centering
\resizebox{\linewidth}{!}{%
\begin{tabular}{l|cccc}
\toprule[1.2pt] 
\textbf{Method}&\textbf{IoU}&\textbf{F$_{1}$ score}&\textbf{AUC}&\textbf{Recall}\\ \hline
{U-Net}                           & 50.61 (0.26) & 65.48 (0.26) & \textbf{96.81} (0.24) & 59.11 (0.58) \\ 
{U-Net w/ Proposed}                & \textbf{52.10} (0.21) & \textbf{66.96} (0.25) & 95.87 (0.19) & \textbf{61.08} (0.60) \\ \hline

{ResUnet}                        & 38.70 (3.71) & 54.80 (3.51) & 95.58 (0.94) & 42.12 (3.84)\\  
{ResUnet w/ Proposed}             & \textbf{41.09} (1.39) & \textbf{56.57} (1.56) & \textbf{96.81} (0.71) & \textbf{46.80} (1.90) \\ \hline
{CE-Net}                         & 37.67 (2.57) & 52.33 (2.89) & 95.26 (0.60) & 38.51 (3.34) \\ 
{CE-Net w/ Proposed}              & \textbf{46.26} (0.86) & \textbf{61.76} (0.89) & \textbf{96.33} (1.47) & \textbf{51.58} (2.68) \\  \hline
{UNet++}                             & 53.30 (0.65) & 67.78 (0.41) & \textbf{97.13} (0.79) & 63.35 (3.42)  \\
{UNet++ w/ $\mathcal{L}_{pd}$}       & 54.24 (0.19) & 68.38 (0.20) & \underline{97.00} (1.08) & \underline{63.93} (1.19)  \\
{UNet++ w/ $\mathcal{L}_{de}$}       & \underline{54.91} (0.36) & \underline{68.85} (0.36) & 96.32 (0.51) & 63.74 (0.77)  \\
{UNet++ w/ Proposed}              & \textbf{55.38} (0.25) & \textbf{69.34} (0.24) & 96.50 (0.08) & \textbf{64.86} (0.74) \\

\bottomrule[1.2pt]
\end{tabular}
}
\caption{Average results of the proposed modules combined with other CNN backbones for hard exudate segmentation. Numbers in parentheses are standard deviations.}
\label{tab:ablation}
\end{table}

\begin{figure}[t]
    \centering
    \includegraphics[width = \linewidth]{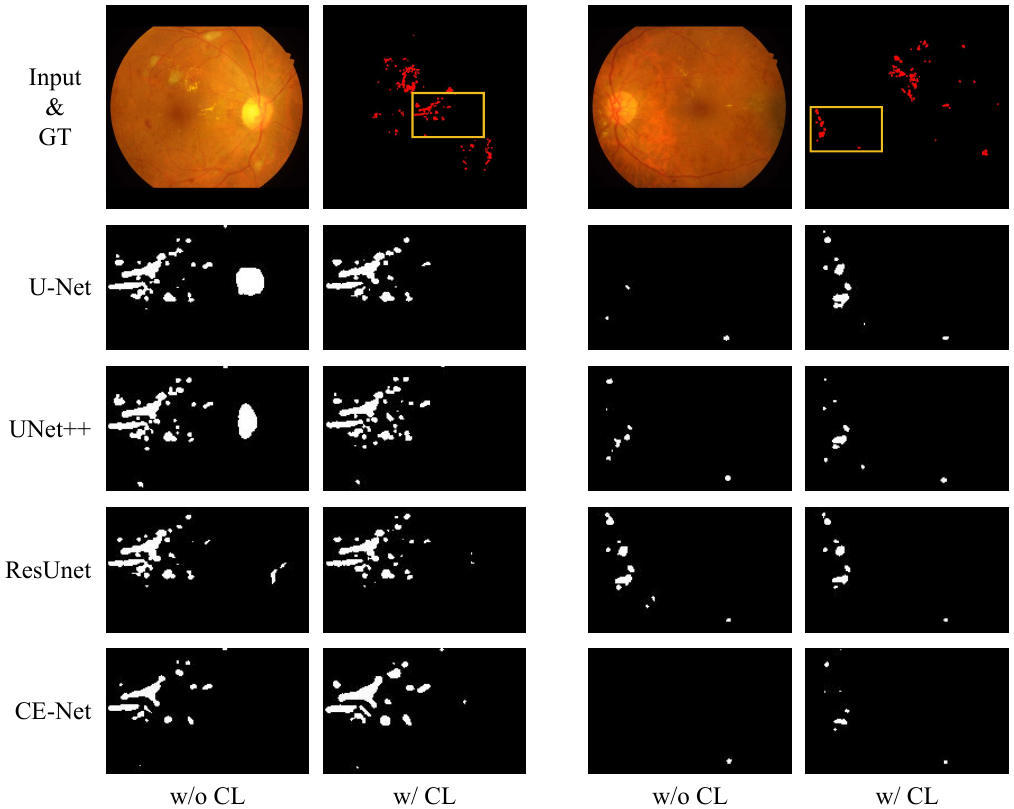}
    \vspace{-0.5cm}
    \caption{Comparison of network segmentations with (w/ CL) and without (w/o CL) the proposed framework. Top row: input images and GT masks. Following rows: segmentation outputs, with yellow boxes emphasizing optic disc and small lesion areas.}
    \label{fig:visual}
    \vspace{-0.5cm}
\end{figure}

Table~\ref{tab:ablation} highlights the enhanced performance of the proposed framework over the vanilla networks, evidenced across three different backbones. The proposed method delivers substantial improvements in all evaluated metrics, indicating a segmentation that is both more precise and reliable. The qualitative evidence, shown in Figure \ref{fig:visual}, further confirms this quantitative assessment. Our method proficiently identifies subtle lesions and ambiguous boundaries, while successfully preventing the misclassification of structures such as the optic disc. Furthermore, our ablation study demonstrates the generalizability of the proposed framework that consistently enhances the performance of various network architectures.

\section{Conclusion and Future Works}
\label{sec:majhead}

The study has shown that integrating the patch-wise density contrasting scheme and discriminative edge inspection module by supervised contrastive losses considerably improves the accuracy of hard exudate detection. The robustness and adaptability of our framework have been further confirmed through comprehensive ablation studies. Future work will extend our framework for segmentation across additional hard exudate datasets with domain shift. Given its generalizability, our proposed method may prove effective for various medical imaging tasks that present challenges similar to those of hard exudate detection.


\section{Compliance with Ethical Standards}
This research study was conducted retrospectively using human subject data made available in open access by~\cite{porwal2018indian}. Ethical approval was not required as confirmed by the license
attached with the open access data.

\section{Acknowledgement}
This work was partially supported by HKRGC GRF grants CityU1101120, CityU11309922, CRF grant C1013-21GF, and HKRGC-NSFC Grant N\_CityU214/19. The authors would like to thank Dr. Jizhou Li and the Hong Kong Centre for Cerebro-Cardiovascular Health Engineering (\href{https://www.hkcoche.org/}{hkcoche.org}) for the collaboration and support in this research.



\small
\bibliographystyle{IEEEbib}
\bibliography{refs}

\end{document}